# Multiband Superconductivity in the time reversal symmetry broken superconductor $Re_6Zr$

Pradnya Parab[a,b], Deepak Singh[c], Harries Muthurajan[b], R. P. Singh[c*], Pratap Raychaudhuri[d†], Sangita Bose[a‡]

[a] *UM-DAE Center for Excellence in Basic Sciences, University of Mumbai, Kalina Campus, Mumbai 400098, India*

[b] *National Centre for Nanoscience & Nanotechnology, University of Mumbai, Kalina Campus, Mumbai 400098, India*

[c] *Department of Physics, Indian Institute of Science Education and Research Bhopal, Bhopal Bypass Road, Bhauri, Bhopal 462066, Madhya Pradesh, India*

[d] *Tata Institute of Fundamental Research, Homi Bhabha Road, Colaba, Mumbai 400005, India*

We report point contact Andreev Reflection (PCAR) measurements on a high-quality single crystal of the non-centrosymmetric superconductor $Re_6Zr$. We observe that the PCAR spectra can be fitted by taking two isotropic superconducting gaps with $\Delta_1 \sim 0.79$ meV and $\Delta_2 \sim 0.22$ meV respectively, suggesting that there are at least two bands which contribute to superconductivity. Combined with the observation of time reversal symmetry breaking at the superconducting transition from muon spin relaxation measurements, our results imply an unconventional superconducting order in this compound: A multiband singlet state that breaks time reversal symmetry or a triplet state dominated by interband pairing.

* rpsingh@iiserb.ac.in
† pratap@tifr.res.in
‡ sangita.bose@gmail.com

In recent times, non-centrosymmetric superconductors (NCS) have attracted considerable interest owing to the complex nature of superconductivity in these materials[1]. In conventional superconductors, where the crystal structure possesses a center of inversion symmetry the superconducting order parameter (OP) is characterized by a distinct parity corresponding to either a spin-singlet or spin-triplet pairing. However, when the crystal structure lacks a point of inversion symmetry, parity and hence spin is no longer a good quantum number and this classification is no longer adequate. In the absence of inversion symmetry, the antisymmetric spin-orbit coupling (ASOC) lifts the degeneracy between the Bloch states with same crystal momentum, $\boldsymbol{k}$, but opposite spins, giving rise to two helicity bands where the Fermion spins are pinned along specific directions with respect to $\boldsymbol{k}$. When the ASOC is large enough that the pairing only occurs between electrons in the same helicity band, the OP can be a mixture of spin-singlet spin-triplet state. If this spin-triplet component is large a NCS can exhibit unconventional properties, such as the upper critical field exceeding the Pauli limit and nodes developing in the superconducting energy gap.

Despite several theoretical predictions, experimental evidence of singlet-triplet mixing in NCS have been surprisingly few. Among NCS that are superconductor at ambient pressure, evidence of nodes in the superconducting energy gap has been obtained for $Li_2Pt_3B$ [2,3], $Y_2C_3$ [4] and $CePt_3Si$ [5,6], though in the last one study of parity broken superconducting state is complicated by coexistence of antiferromagnetic order. Yet many other compounds, such as BiPd [7,8,9,10], $Nb_{0.18}Re_{0.82}$ [11,12], $Re_3W$ [13], $PbTaSe_2$ [14] exhibit fully open superconducting gap while some of them show evidence for multiband superconductivity. Recently, the superconducting state of the NCS, $Re_6Zr$ with the α-Mn cubic structure[15] has called for particular attention. While specific heat, penetration depth and nuclear quadrupole resonance (NQR) data are consistent with a fully gapped state, the upper critical field ($H_{c2}$) is very close to the Pauli limit[16,17].

More importantly muon spin rotation (μSR) measurements performed on polycrystalline samples reveal that the superconducting state breaks time reversal symmetry (TRS), which the authors suggest as evidence for strong spin singlet-spin triplet mixing[18]. However, the same measurement provides evidence of a Bardeen-Cooper-Schriffer (BCS) like variation of superfluid density, consistent with a fully gapped s-wave state. Since spin singlet-triplet mixing is expected to give a rise to strong anisotropy in the gap function, it is important to obtain detailed information of the superconducting gap symmetry on high quality single crystals.

Point contact Andreev reflection (PCAR) spectroscopy[19] is a powerful tool to investigate the gap symmetry in superconductors. In this technique, a ballistic contact is established by bringing a sharp normal metal tip in contact with the superconductor. The dependence of the differential conductance ($G(V) = dI/dV$) as a function bias voltage ($V$) of such a contact is sensitive to the magnitude and symmetry of the superconducting gap function. Consequently, $G(V)$-$V$ spectra provides valuable insight on the gap symmetry and its temperature evolution in unconventional superconductors.

In this paper, we report PCAR measurements on a high quality single crystal of $Re_6Zr$.The single crystal was grown using Czochralski crystal pulling method using a tetra-arc furnace under argon atmosphere, starting from a polycrystalline $Re_6Zr$ button. Polycrystalline sample of $Re_6Zr$ was prepared by arc melting stoichiometric quantities of Re (99.99%, Alfa Aesar) and Zr (99.99%, Alfa Aesar) on a water-cooled copper hearth in a high-purity Ar atmosphere in tetra-arc furnace. The button was melted several time to ensure phase homogeneity. The observed weight loss during the melting was negligible. The phase purity of the polycrystalline button was checked by powder X-ray diffraction. In the Czochralski growth, a tungsten rod was used as the seed, and the crystal was pulled at the rate of 30 – 50 mm/h. The Laue diffraction measurement performed on the crystal revealed sharp Bragg spots consistent with growth along the principal axis of the crystal (inset Fig. 1(a)). Figure 1(a)-(b) shows the basic characterization of superconducting properties from in-field transport measurements. In zero field (Fig. 1(a)) we observe a sharp superconducting transition with $T_c$ ~ 6.8 K and a normal state resistivity $\rho$ ~ 200 μm-cm at 10 K. The temperature variation of upper critical field ($H_{c2}$) is determined from the temperature at which the resistivity is less than 1% of the normal state resistivity in fixed magnetic field $\rho$ - T scans. At 1.4 K, $H_{c2}$ ~ 110 kOe consistent with earlier reports on polycrystalline samples[16].

For PCAR measurements we establish ballistic contacts by mechanically engaging a fine Ag tip on [100] surface of the $Re_6Zr$ single crystal inside a conventional $^4$He cryostat (Fig. 2(a)). The differential conductance, $G(V)$, is obtained by numerically differentiating the current versus voltage ($I$-$V$) characteristics recorded at a fixed temperature. In Figure 2(b)-(e) we show representative $G(V)$-$V$ spectra for four different contacts recorded at temperatures below 2 K. All the spectra display two symmetric coherence peaks associated with the superconducting energy gap and a dip at zero bias characteristic of an s-wave superconductor. In addition to this dominant feature the spectrum in Fig. 2(e) also shows small additional weak feature close to 0.3 meV signifying the possible existence of a second gap. This feature is however delicate: Slightly elevated temperature or small amount of broadening smears this feature making it unresolvable[20]. To obtain quantitative information we first attempt to fit the spectra with the Blonder-Tinkham-Kalpwijk[21] (BTK) model with an isotropic gap, generalized to take into account broadening effects[22]. We use the superconducting energy gap (Δ), the dimensionless barrier potential at the interface (Z) and the phenomenological broadening parameter (Γ) as fitting parameters. While Γ is formally introduced to account for the finite life of the quasiparticles, in practice this parameter phenomenologically incorporates all non-thermal sources of broadening, e.g. distribution of superconducting energy gaps, broadening due to interface defects as well as instrumentational broadening. We observe that while the fits broadly capture the shape of the spectra, there are significant deviations. First, all the fits deviate significantly at bias voltages above the coherence peak. Secondly, for the spectrum in Fig. 2(e) the fit does not account for the feature seen close to 0.3 meV. Next, we investigate whether the difference could arise from multiband superconductivity, where different superconducting gaps could co-exist in the same superconductor. We use a two-band BKT model[23], where, $I$, and hence $G$ is a weighted sum from two transport channels [$G_1(V)$ and $G_2(V)$], arising from two bands in the superconductor. In this model the normalized conductance $G(V)/G_N$ (where $G_N = G(V)/G(V \gg \Delta/e)$, where $e$ is the electronic charge) is given by, $G(V)/G_N = (1\text{-}w)G_1(V)/G_{N1} + wG_2(V)/G_{N2}$. $G_1(V)/G_{1N}$ and $G_2(V)/G_{2N}$ are calculated using the generalized BTK formalism using the relative weight factors of the two gaps ($w$), superconducting energy gaps ($\Delta_1$ and $\Delta_2$), the barrier potentials ($Z_1$ and $Z_2$), and the broadening parameters ($\Gamma_1$ and $\Gamma_2$) as fitting parameters. We observe that the two-gap model with two isotropic gaps provides a good fit for all the spectra. We obtain a large gap, $\Delta_1 \sim 0.77 \pm 0.03$ meV, and a small gap, $\Delta_2 \sim 0.23 \pm 0.03$ meV and $w \sim 0.35$ - $0.52$. $\Gamma_1$, $\Gamma_2 < 0.1$ meV, but vary across contacts. We believe that the primary source of spectral broadening here is from defects at the interface between the sample and the tip which varies from contact to contact. On the other hand, additional refinements, such as incorporating anisotropic gap function does not produce any significant improvement in the fit. It is also important to note that in none of the spectra we observe any evidence of any zero bias conductance peak that is expected to arise when the superconducting order parameter changes sign over the Fermi Surface.

We now investigate the temperature dependence of the two gap by tracking the temperature dependence of the point contact shown in Fig. 2(c). We record the $G(V)$-$V$ spectra at different temperatures and fit with the two gap BTK model (Fig. 3(a)). As expected $Z_1$ and $Z_2$ show a small variation over the entire temperature range, whereas *w* is temperature independent (inset Fig. 3(c)). The temperature variation of $\Delta_1$ and $\Delta_2$ are shown in the main panel of Fig. 3(b). $\Delta_1$ follows the expected temperature dependence from Bardeen-Cooper-Schriffer (BCS) theory at low temperature and marginally deviates towards lower values close to $T_c$. $\Delta_2$ on the hand also follows BCS-like variation at low temperatures, but with a lower $T_c$ of 4.6 K, and forms a tail above this temperatures. (Above 5K, we cannot resolve $\Delta_2$ from our data.) Both these features are characteristic of a two-band superconductor with small, but finite interband scattering. $\Gamma_1$ and $\Gamma_2$ are both small (<0.01 meV) and marginally increase close to $T_c$ (Fig. 3(c)).

We now discuss the implication of these results. First we explore whether the two gap feature observed in our PCAR experiments could arise from singlet-triplet mixing. For a NCS, ASOC leads to a term of the form $\alpha g(\boldsymbol{k})\cdot\sigma$ in the Hamiltonian, where α is the spin-orbit coupling constant, σ is the Pauli matrices, and the vector $g(\boldsymbol{k})$, representing the orbital direction, obeys the antisymmetric property such that $g(\boldsymbol{k}) = -g(-\boldsymbol{k})$. The ASOC breaks the spin degeneracy, which leads to two bands characterized by ± helicities for which the spin eigenstates are either parallel or antiparallel to $g(\boldsymbol{k})$. The superconducting energy gap on both these bands will have a singlet and a triplet component. The system will therefore behave like a two-band superconductor with two gaps defined on each of ASOC split bands, $\Delta_{\pm}(\bar{k})=\Delta_s \pm \Delta_t(\bar{k})$, where $\Delta_s$ and $\Delta_t(\boldsymbol{k})$ are the spin singlet and spin triplet component of the gap function. Since $\Delta_t(\boldsymbol{k})$ is strongly anisotropic and changes sign depending on $\boldsymbol{k}$ direction, a significant $\Delta_t(\boldsymbol{k})$ component implies that both $\Delta_{\pm}(\bar{k})$ would be strongly anisotropic with a large distribution of gap amplitude over the Fermi surface. Since in a PCAR experiment the current through the contact has contribution from all over the Fermi Surface, it has been shown that such a gap distribution translates into systematic large values of the broadening parameter[24,20] , Γ. In contrast, in our experiment we can fit the many spectra with very small values of $\Gamma_1/\Delta_1$ and $\Gamma_2/\Delta_2$, suggesting that the anisotropy of both gaps is very small. We therefore conclude that spin-singlet spin-triplet mixing if at all present is very small and cannot account for the two gap that we observe. This is also expected from the small band splitting due to ASOC (~30 meV)[16] calculated for this compound which is comparable to the ASOC spin splitting in $Li_2Pd_3B$ where a fully gapped isotropic order parameter has been inferred from penetration depth measurements[25]. (In the isostructural NSC $Li_2Pt_3B$ where penetration depth measurements provide evidence of large anisotropy and possible nodes in the gap function, the ASOC band splitting is ~200 meV.) On the other hand, our results are completely consistent with conventional multiband superconductivity which arise when two isotropic energy gaps are present on two different Fermi surface pockets. In that case, the salient question is, *how can one reconcile these results with the observation TRS breaking in μSR measurements?*

There are two possible scenario where a multiband superconductor with fully open superconducting gap can break TRS. The *first scenario* has been proposed for a multiband superconductor with high symmetry[26] such as the cubic structure of $Re_6Zr$. In such a system, TRS can get broken even with conventional s-wave singlet pairing. Under certain condition, when the Coulomb repulsion between two Fermi surface pockets dominate, the relative phase between one pocket and another will be non-zero. Such a superconducting state will break TRS, allowing antiferromagnetic domains and fractional vortices to appear. A *second scenario* has been suggested in the context of TRS breaking in $LaNiC_2$ and $LaNiGa_2$ [27]. Here again the system consists of two bands, but superconductivity is dominated by interband pairing between the two bands. When the pairing is antisymmetric with respect of interchange of the band index, TRS breaking triplet superconductivity can be realized even when the Fermi surface remains fully gapped. One would get two gaps of different sizes corresponding to $\uparrow\uparrow$ and $\downarrow\downarrow$ pairing. Distinguishing between these two scenarios would require a detailed knowledge of the Fermi surface structure properties in this compound. Experimentally, it might be possible to distinguish these two scenarios from PCAR measurement using a highly spin polarized tip.

In summary, from a detailed analysis of PCAR spectra on a $Re_6Zr$ single crystal we have presented evidence of multiband superconductivity in this compound. Our measurement suggest that the superconducting state is composed of two fairly isotropic gap functions, which rule out the possibility of spin-singlet spin-triplet mixing as the origin of the multiband behavior. On the other hand it is more likely that the two-gap are associated with the existence of multiple Fermi surface sheets. Combined with μSR measurements which find that the TRS is broken at the superconducting transition, our observation point towards an unusual superconducting state: Either a multiband singlet state that breaks TRS or a triplet state driven by interband pairing. Validation of either of these two scenarios will require further theoretical and experimental studies.

*Acknowledgements:* The authors would like to thank Vivas Bagwe and John Jesudasan for technical help and Daniel Agterberg for critically reading the manuscript. The work was supported by Department of Atomic Energy, Government of India, Department of Science and Technology, Government of India (Grant No: EMR/2015/000083, SERB/F/1877/2012 and SERB/F/745/2014) and Indian National Science Academy (Grant No: SP/YSP/73/2012/1875).

---

[24] P. Raychaudhuri, D. Jaiswal-Nagar, G. Sheet, S. Ramakrishnan and H. Takeya, *Evidence of gap anisotropy in superconducting* $YNi_2B_2C$ *using directional point contact spectroscopy.* Phys. Rev. Lett. **93,** 156802 (2004)

[25] K.-W. Lee and W. E. Pickett, *Crystal symmetry, electron-phonon coupling, and superconducting tendencies in* $Li_2Pd_3B$ *and* $Li_2Pt_3B$. Phys. Rev. B **72**, 174505 (2005).

[26] D. F. Agterberg, V. Barzykin, and L. P. Gor'kov, *Conventional mechanisms for exotic superconductivity.* Phys. Rev. B **60**, 14868 (1999).

[27] Z. F. Weng, J. L. Zhang, M. Smidman, T. Shang, J. Quintanilla, J. F. Annett, M. Nicklas, G. M. Pang, L. Jiao, W. B. Jiang, Y. Chen, F. Steglich, and H. Q. Yuan, *Two-Gap Superconductivity in* $LaNiGa_2$ *with Nonunitary Triplet Pairing and Even Parity Gap Symmetry.* Phys. Rev. Lett. 117, 027001 (2016).

## Figure Captions

**Figure 1.** (a) Resistivity ($\rho$) vs Temperature (T) for $Re_6Zr$ single crystal in zero field showing sharp superconducting transition with Tc ~ 6.8 K. The inset shows X-ray Laue diffraction pattern of crystal along the principal axis of the crystal. (b) Temperature dependence of $H_{c2}$. The inset shows $\rho$-T plot at various magnetic field.

**Figure 2.** (a) Schematic of the point contact between $Re_6Zr$ sample and an Ag tip. (b)-(e) Representative PCAR *G(V)* vs *V* spectra recorded below 2K for different contacts on the [100] surface of $Re_6Zr$ single crystal. The red solid line shows the fit using the two gap BTK model. The black solid line shows the best fit using the one-gap model. The values of the best fit parameters are shown in the respective panels.

**Figure 3.** (a) Temperature evolution of the PCAR *G(V)* vs *V* spectra for the contact shown in Fig 2(c); for clarity, each successive plot other than the one at 6.5 K are vertically scaled up by multiplying by a factor of (1+0.1(*n*-1)), where *n* stands for the serial number of the plot from bottom. The solid lines are the fit to the two-gap BTK model. (b) Temperature variation $\Delta_1$ and $\Delta_2$. The solid black line shows the expected BCS temperature for $\Delta_1$. The solid grey line shows the expected BCS variation for $\Delta_2$ assuming a lower $T_c$ of 4.6 K. (c) Temperature variation of barrier potential ($Z_1$ and $Z_2$) and broadening parameters ($\Gamma_1$ and $\Gamma_2$) obtained from best fit data. The inset shows the relative weight factor, *w*, as a function of temperature.

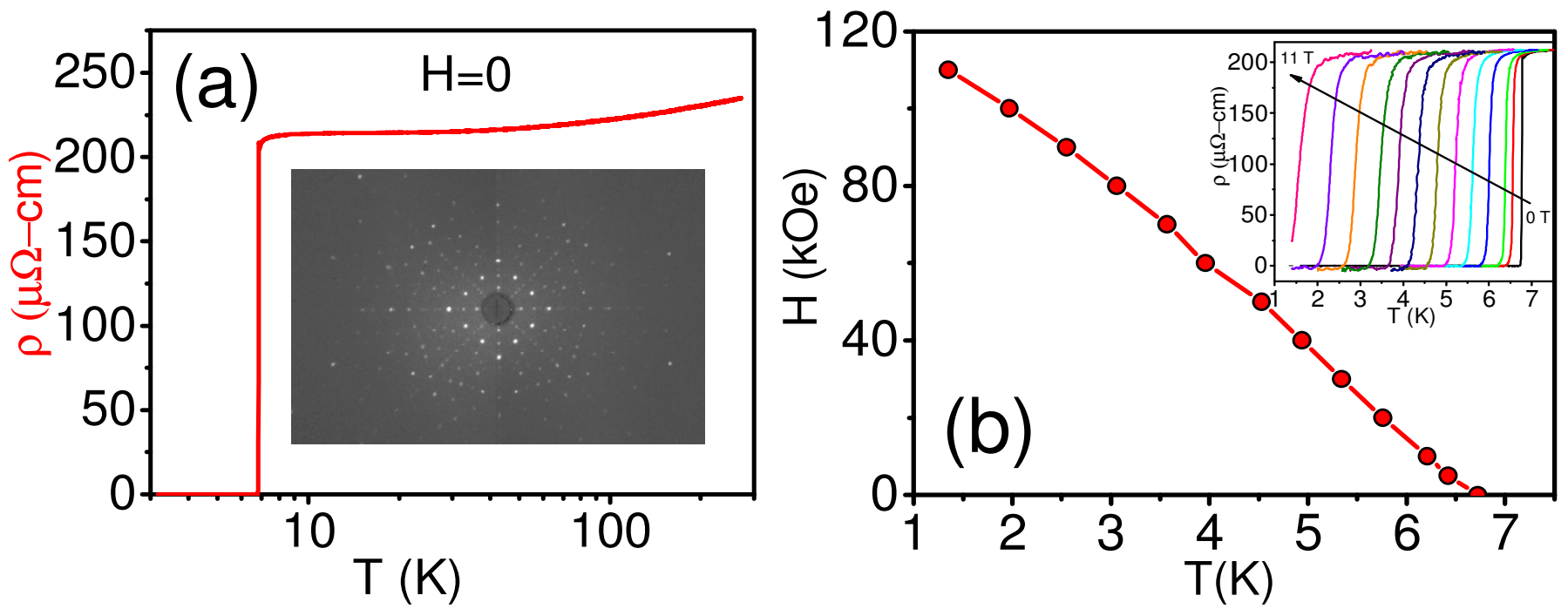

**Figure 1**

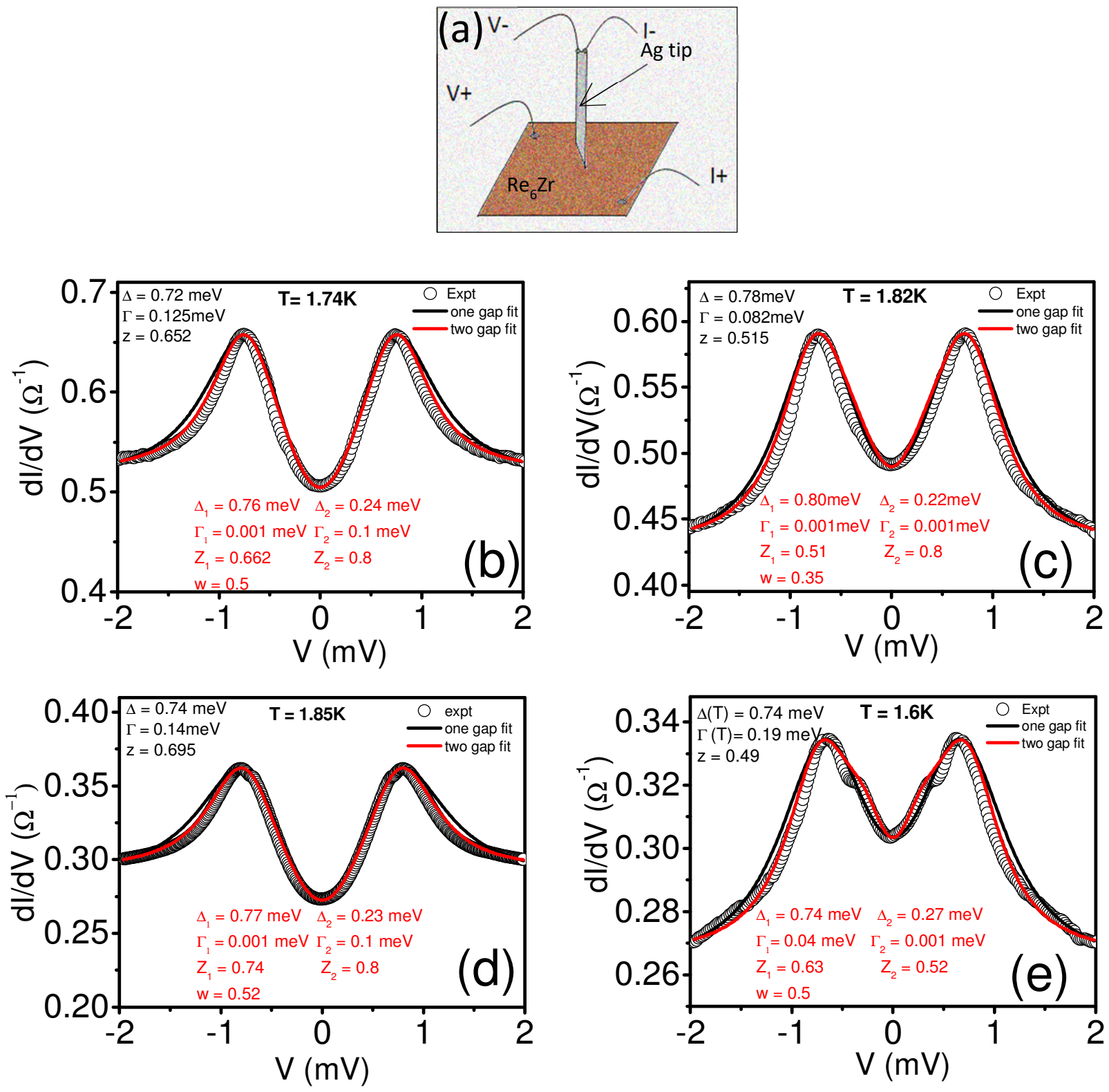
(a)
V-
I-
Ag tip
V+
Re6Zr
I+
Δ = 0.72 meV
Γ = 0.125meV
z = 0.652
T= 1.74K
Expt
one gap fit
two gap fit
Δ1 = 0.76 meV Δ2 = 0.24 meV
Γ1 = 0.001 meV Γ2 = 0.1 meV
Z1 = 0.662 Z2 = 0.8
w = 0.5
(b)
dI/dV (Ω-1)
V (mV)
Δ = 0.78meV
Γ = 0.082meV
z = 0.515
T = 1.82K
Expt
one gap fit
two gap fit
Δ1 = 0.80meV Δ2 = 0.22meV
Γ1 = 0.001meV Γ2 = 0.001meV
Z1 = 0.51 Z2 = 0.8
w = 0.35
(c)
dI/dV(Ω-1)
V (mV)
Δ = 0.74 meV
Γ = 0.14meV
z = 0.695
T = 1.85K
expt
one gap fit
two gap fit
Δ1 = 0.77 meV Δ2 = 0.23 meV
Γ1 = 0.001 meV Γ2 = 0.1 meV
Z1 = 0.74 Z2 = 0.8
w = 0.52
(d)
dI/dV (Ω-1)
V (mV)
Δ(T) = 0.74 meV
Γ (T)= 0.19 meV
z = 0.49
T = 1.6K
expt
one gap fit
two gap fit
Δ1 = 0.74 meV Δ2 = 0.27 meV
Γ1 = 0.04 meV Γ2 = 0.001 meV
Z1 = 0.63 Z2 = 0.52
w = 0.5
(e)
dI/dV (Ω-1)
V (mV)

**Figure 2**

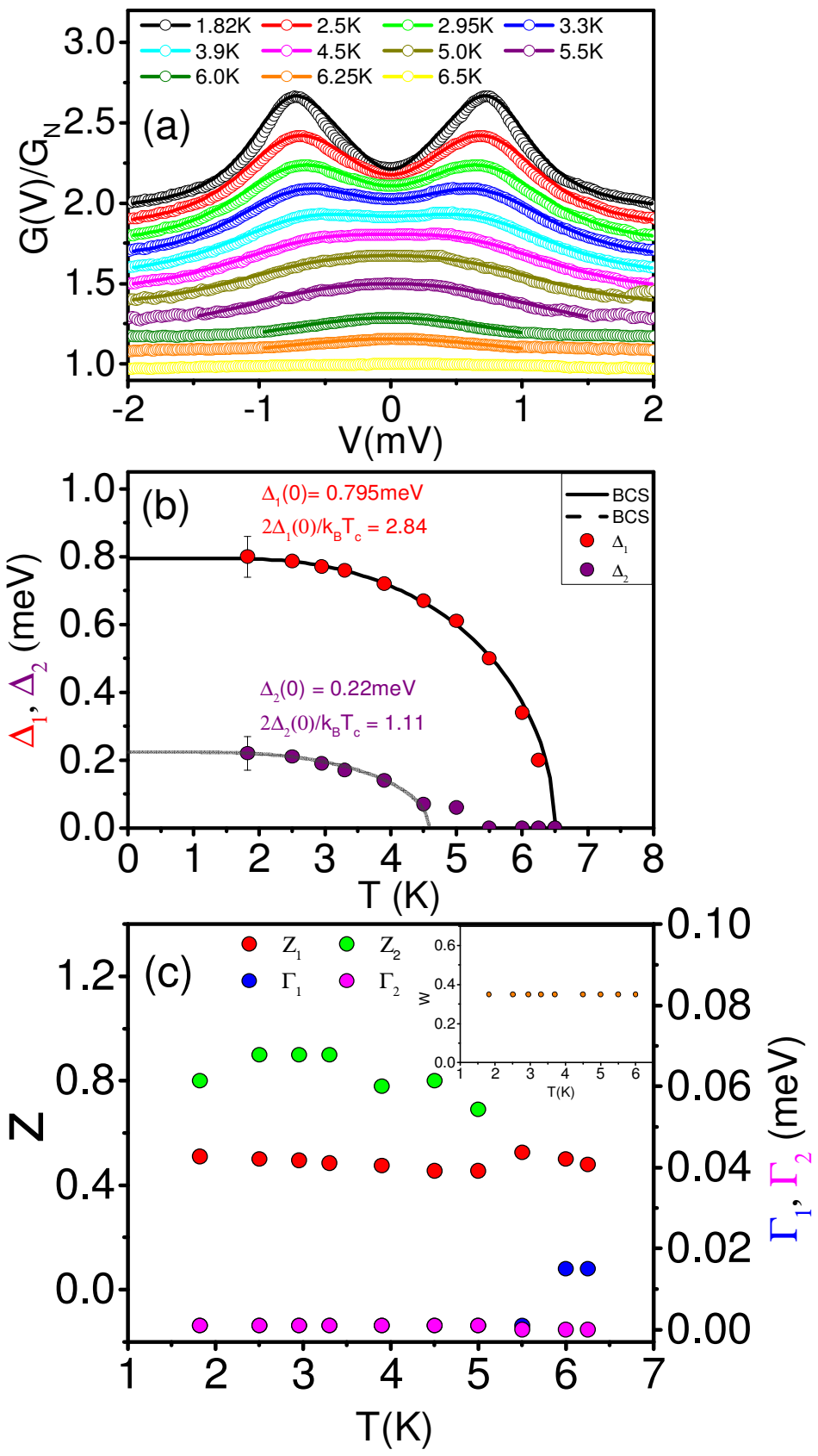

**Figure 3**